\documentclass[12pt,epsf]{article}
\usepackage{epsfig}
\usepackage{amssymb}  
\usepackage{amsthm}
\usepackage{amsmath}
\let\eqref\cref

\renewcommand{\thanks}[1]{\footnote{#1}} 

\renewcommand{\theequation}{\thesection.\arabic{equation}}
\newcommand{\be}{\begin{equation}}
\newcommand{\ee}{\end{equation}}
\newcommand{\bea}{\begin{eqnarray}}
\newcommand{\eea}{\end{eqnarray}}

\def\stacksymbols#1#2#3#4{\def\theguybelow{#2}
\def\verticalposition{\lower#3pt}
\def\spacingwithinsymbol{\baselineskip0pt\lineskip#4pt}
\mathrel{\mathpalette\intermediary#1}}
\def\intermediary#1#2{\verticalposition\vbox{\spacingwithinsymbol
\everycr={}\tabskip0pt
\halign{$\mathsurround0pt#1\hfil##\hfil$\crcr#2\crcr
\theguybelow\crcr}}}

\newcommand{\gapproxeq}{\stacksymbols{>}{\sim}{3}{.5}}

\begin{document}

\pagestyle{empty}

\bigskip\bigskip
\begin{center}
{\bf \large Radiations From Geometries Using Painlev\'e-Gullstrand Coordinates}
\end{center}

\begin{center}
Pradip Kattel\footnote{e-mail address, pradip.kattel@bison.howard.edu} and
James Lindesay\footnote{e-mail address, jllindesay@howard.edu} \\
Computational Physics Laboratory \\
Howard University,
Washington, D.C. 20059 
\end{center}
\bigskip
\begin{center}
{\bf Abstract} 
\end{center}
The principle of equivalence is used to examine covariant
descriptions of quantum phenomena within the global exterior
of geometries described using Painlev\'e-Gullstrand coordinates,
which are everywhere non-singular away from their center. 
The differences between the descriptions of
observers  stationary within the curvilinear geometry 
and those in a locally-flat
freely-falling system that is momentarily
at rest are found to depend \emph{only}
on local proper accelerations, becoming vanishingly small
for increasingly distant observers.  Away from the horizon,
the local detections of outgoing massless radiations from the
ground state of the freely-falling observers are found
to be consistent with those of Rindler observers accelerating in
flat space-time.

\bigskip \bigskip 

\setcounter{equation}{0}
\section{Introduction}
\indent \indent
When exploring the subtleties of merging quantum mechanics with the classical theory of general relativity, it is often convenient to explore
the implementation of quantum principles into stationary spherically symmetric geometries using the principle of equivalence.  
The Schwarzschild metric represents a \emph{global} vacuum solution to Einstein's field equations, assuming a diagonal metric form. 
This metric introduces a time coordinate that has a non-physical coordinate singularity.  In contrast, the
Painlev\'e-Gullstrand metric\cite{PGmetric, rivermodel} likewise represents a \emph{global} vacuum solution that relaxes the requirement of a diagonal
metric form, thereby introducing no non-physical coordinate singularities, making these coordinate convenient for describing the
quantum dynamics of energy-momenta conjugate to these coordinates.  Although the two metric forms can be \emph{locally} related
away from coordinate singularities, no coordinate transformation that is everywhere non-singular can relate these metrics. 
In particular, the use of the Schwarzschild time coordinates near a black hole horizon as the temporal coordinate describing
the space-time dynamics transforms that coordinate singularity into a physical singularity in Einstein's equation.

In what follows, we will implement local quantum principles in a global Painlev\'e-Gullstrand geometry via
the principle of equivalence to examine geometric effects on local physical measurements, as well as 
determine the limits of applicability of local flatness.
Furthermore, we will compare and contrast descriptions of physical
dynamics described using Painlev\'e-Gullstrand coordinates with Rindler coordinates and
Schwarzschild coordinates in $1+1$ dimensional space-time, particularly with regards to outgoing massless radiations. 


\section{Stationary Spherically-Symmetric Geometries and the Principle of Equivalence}
\indent \indent
The global vacuum solution to Einstein's equations for stationary, spherically-symmetric systems that will here be
utilized has a metric of the form
\begin{equation}\label{RivInterval}
ds^2 = -\left(1-\frac{R_S}{r}\right) dct^2 + 2 \sqrt{\frac{R_S}{r}} dct \, dr+ dr^2 + r^2 (d \vartheta^2 + \sin^2 \vartheta d \varphi^2),
\end{equation}
where $R_S \equiv \frac{2 G_N \, M}{c^2}$ is the Schwarzschild radius associated with a geometry generated
by mass $M$.  We will in what follows examine only the 1+1 dimensional system absent any variations in the angular coordinates. 
The global coordinates $(ct,r)$ manifest no coordinate singularities away from the physical singularity
at $r=0$ in this classical geometry, allowing their direct use in describing local dynamics everywhere. 
The metric exhibits an off-diagonal $dct \, dr$ term expected
for radially dynamic (i.e., accreting or excreting) geometries as an analog to the $dct \, d\varphi$ term in the
rotationally dynamic Kerr metric \cite{KerrGeometry}.

There has yet to be demonstrated any gravitating physical phenomena, either classical or quantum in nature,
violating the principle of equivalence.  In contrast, several generally accepted classical and quantum experiments
have demonstrated the viability of this principle\cite{JLFQGreferences}.  Thus, we consider the
principle of equivalence to be an established foundational paradigm in this paper. 
For implementation of the principle of equivalence, a set of freely-falling coordinates will next be generated.


\subsection*{Radially freely-falling coordinates}
\indent \indent
We next develop a global set of freely-falling coordinates for the 1+1 dimensional geometry with metric
\begin{equation}
ds_{1+1}^2 = -\left(1-\frac{R_S}{r}\right) dct^2 + 2 \sqrt{\frac{R_S}{r}} dct \, dr+ dr^2  .
\label{RivIntervalctr}
\end{equation}
Geodesic motion in the $(ct,r)$ coordinates have a solution set satisfying \\
$\left \{ \frac{d ct(c\tau)}{d c\tau}=1,
\frac{d r(c\tau)}{d c\tau}=-\sqrt{\frac{R_s}{r(c\tau)}} \right \}$
in terms of the proper time $\tau$ parameterizing
the freely-falling observers $r(c\tau)$.
Choosing the constant of integration $\rho$ as a static space-like coordinate associated with
the freely-falling system, one can develop new \emph{global} coordinates given by
\begin{equation}
ct=\omega, \ \  r= \left(\frac{3}{2}\right)^\frac{2}{3} \left(R_S (\rho -\omega)^2 \right)^\frac{1}{3} .
\label{Eq:LamaitreJL}
\end{equation}
These coordinates are similar to the Lama\^itre coordinates\cite{Lamaitre}
developed from Schwarzschild space-time. 
Using these coordinates, a particular free-faller at fixed coordinate $\rho_p$ reaches the singular
center at time $\omega=\rho_p$. 
The space-time metric (\ref{RivInterval}) takes the diagonal form
\begin{equation}
\left( \left( \bar{g}_{\alpha\beta} \right) \right) = 
\begin{pmatrix}
-1 & 0 \\
0 & \frac{R_S}{r}
\end{pmatrix} \underset{r\longrightarrow r(\omega, p)}\rightarrow \begin{pmatrix}
-1 & 0 \\
0 & \left(\frac{2}{3}\right)^\frac{2}{3} \left(\frac{R_S}{\rho - \omega} \right)^\frac{2}{3} 
\end{pmatrix} .
\label{omgrhometric}
\end{equation}
It is clear that this global metric form is non-singular
everywhere away from the center $r(\omega,\rho)=0$. 
In 1+1 dimensions, although the Einstein tensor $G_{\mu \nu}$ vanishes everywhere in this
vacuum solution, the curvature scalar $R = - \frac{8}{9(\rho-\omega)^2} \underset{r \rightarrow r(\omega, p)} \Longrightarrow - \frac{2R_S}{r^3}$
does not, making it a convenient measure of local tidal scales.  
The geodesic equations for four-velocities $\left( U^\omega \equiv \frac{d \omega}{d \lambda}, U^\rho \equiv \frac{d \rho}{d \lambda} \right )$
are as follows:
\begin{equation}
\frac{dU^\omega}{d\lambda} +\frac{\left(\frac{2}{3} \right)^\frac{2}{3} R_S^\frac{2}{3} \, (U^\rho)^2}{3(\rho-\omega)^\frac{5}{3}} = 0
\quad , \quad
\frac{dU^\rho}{d\lambda} + \frac{2 U^\omega U^\rho- (U^\rho)^2}{3(\rho-\omega)}  = 0.
\end{equation}
It is clear that if $U^\rho = 0$ for $\rho \rightarrow \rho_p$ the system is not accelerating.


\subsection*{Locally flat coordinates}
\indent \indent
Next, coordinates that are locally flat for the freely falling system at $(\omega_p, \rho_p)$
will be developed for the metric ($\ref{omgrhometric}$).  This \emph{local} set of coordinates
described as $(Y^0,Y^1)$ has the form
\bea
\omega_{\text{flat}}(Y^0,Y^1)= Y^0+ \frac{(Y^1-\omega_p)^2}{6(\omega_p-\rho_p)},  \qquad \qquad  \qquad \qquad \quad \:  \nonumber \\
\rho_\text{flat}(Y^0,Y^1)= \rho_p+ 
\left( \frac{3}{2}\right)^\frac{1}{3} \left( \frac{\rho_p-\omega_p}{R_S}\right)^\frac{1}{3} (Y^1-\rho_p) + \quad  \label{Eq:omgrhoYYFFF} \\
 \frac{(Y^1-\rho_p)^2}{2^\frac{5}{3}3^\frac{1}{2}\left(R_S^2 (\rho_p-\omega_p) \right)^\frac{1}{3}} 
- \frac{(Y^1-\rho_p)(Y^0-\omega_p)}{2^\frac{1}{3}3^\frac{2}{3}\left(R_S(\rho_p-\omega_p)^2 \right)^\frac{1}{3}} .    \nonumber         
\eea
One can directly verify that the metric generated by these coordinates satisfies the vacuum Einstein equation with curvature
scalar $R\underset{(Y^0,Y^1) \rightarrow (\omega_p,\rho_p)} \Longrightarrow -\frac{2R_S}{r_p^3}$, demonstrating only
\emph{local} flatness.  Expressing the accelerating coordinates $(ct,r)$ in terms of $(Y^0,Y^1)$ results in locally-flat
freely-falling coordinates that are in relative motion to the $(ct_p,r_p)$ frame:
\bea
ct(Y^0,Y^1) \simeq
Y^0 - \sqrt{\frac{R_S}{r_p}} \frac{(Y^1-\rho_p)^2}{4 r_p} , \qquad \qquad \qquad \:  \nonumber \\
r(Y^0,Y^1) \simeq
 r_p +  (Y^1-\rho_p)  - \sqrt{\frac{R_S}{r_p}}(Y^0-\omega_p)  + \quad  \label{Eq:ctrYYFFF} \\ 
\frac{R_S}{4 r_p^2} \left [ (Y^1-\rho_p)^2-(Y^0-\omega_p)^2 \right ].   \nonumber
\eea
The expressions are valid as long as coordinate deviations away from the local
point $(\omega_p,\rho_p)$ are small compared to length scales associated with tidal effects,
$(Y^0-\omega_p) << \sqrt{\frac{r_p^3}{2 R_S}}$ and $(Y^1-\rho_p) << \sqrt{\frac{r_p^3}{2 R_S}}$.
These forms can be locally inverted to determine the following useful
local relations for the point $(ct_p,r_p)$:
\be
\frac{\partial Y^0}{\partial ct}=1, \,
\frac{\partial Y^0}{\partial r}=0, \,
\frac{\partial Y^1}{\partial ct}=\sqrt{\frac{R_S}{r_p}}, \,
\frac{\partial Y^1}{\partial r}=1, \,
\textnormal{and } 
\frac{\partial^2 Y^1}{\partial ct^2}=\frac{R_S}{2 r_p^2}.
\label{Eq:Yderivs}
\ee
From these expressions, it is clear that the accelerating point $(ct_p,r_p)$ is moving with
radial Lorentz velocity $\sqrt{\frac{R_S}{r_p}}$ relative to the locally flat freely-falling
coordinate frame $(Y^0,Y^1)$, allowing direct determination of locally freely-falling
coordinates that are momentarily at rest with $(ct_p,r_p)$, if desired.
The locally flat metric using these (moving) coordinates takes the form
\bea
g_{Y^0Y^0} \simeq -1+\frac{R_S(Y^1-\rho_p)^2}{4 r_p^3} , \qquad \qquad \qquad  \qquad \qquad \qquad \qquad \nonumber \\
g_{Y^0Y^1}=g_{Y^1Y^0} \: \simeq \:
\frac{1}{8 r_p^3} \sqrt{\frac{R_S}{r_p}} \: \left [ \: (3r_p-R_S)(Y^1-\rho_p)^2 \: + \right . \qquad \quad \nonumber \\
-6\sqrt{r_p R_S}(Y^1-\rho_p)(Y^0-\omega_p)+ 
\left . 4R_S(Y^0-\omega_p)^2 \: \right ] ,  \qquad \\
g_{Y^1Y^1} \simeq 1+ \frac{R_S}{4 r_p^3} \left[ (Y^1-\rho_p)^2  -
2 \sqrt{\frac{R_S}{r_p}} (Y^1-\rho_p)(Y^0-\omega_p) \right ] .  \: \nonumber
\eea
Thus, coordinates $(Y^0,Y^1)$ still generate the appropriate gravitational tidal forces
(along with any local energy densities, in general) due to the metric's quadratic form. 
These coordinates will be quite useful for describing local measurements implemented using the
principle of equivalence in the discussion that follows.


\section{Massless Radiations Examined by Inertial and Accelerating Observers}
\indent \indent
We next develop descriptions of massless particles in the geometry described by (\ref{RivIntervalctr})
in exterior regions $r>R_S$. 
The metric can be factored in the form
\be
ds_{1+1}^2 =\left[ \: dr - \left( 1 - \sqrt{ \frac{R_S}{r} } \right) \, dct \: \right ]
\left[ \: dr + \left( 1 + \sqrt{ \frac{R_S}{r} } \right) \, dct \: \right] .
\ee
Light-like motions can be described using null geodesics defined by
\be
 \left( 1 - \sqrt{ \frac{R_S}{r} } \right) \left[ \: dct - dr_{\tilde{u}}  \: \right] = 0 \quad \textnormal{and} \quad
 \left( 1 + \sqrt{ \frac{R_S}{r} } \right) \left[ \: dct + dr_{\tilde{v}} \: \right] = 0,
\label{Eq:NullForms}
\ee
where 
\bea
r_{\tilde{u}}(r) \equiv r+2 \sqrt{r R_S} + 2 R_S \log\left( \sqrt{\frac{r}{R_S}} - 1 \right),  \nonumber \\
r_{\tilde{v}}(r) \equiv r - 2 \sqrt{r R_S} + 2 R_S \log\left( \sqrt{\frac{r}{R_S}} + 1 \right).
\label{Eq:rurv}
\eea
Conformal coordinates $\tilde{u}$ and $\tilde{v}$ parameterizing distinct
outgoing and ingoing light-like trajectories will be defined using (\ref{Eq:NullForms}) in the forms
\be
\tilde{u} \equiv c t- r_{\tilde{u}} \quad \textnormal{and} \quad \tilde{v} \equiv ct+r_{\tilde{v}} .
\label{Eq:uFvF}
\ee
Although the analytic form of $\tilde{v}-\tilde{u}$ is identical
to that from the Schwarzschild metric, the form of $\tilde{v}+\tilde{u}$
is drastically different due to the non-singular near-horizon
behavior of the time parameter $ct$. 
The geodesic equations satisfied by the four-velocities
$U^\beta \equiv \frac{d \tilde{x}^\beta}{d \lambda}$ of these coordinates are given by
\be
\frac{dU^{\tilde{u}} }{d\lambda} - \frac{R_S }{2 r^2 } \left( U^{\tilde{u}} \right)^2= 0 \quad \textnormal{and} \quad
\frac{dU^{\tilde{v}} }{d\lambda} + \frac{R_S }{2 r^2 } \left( U^{\tilde{v}} \right)^2= 0,
\ee
verifying that constant $\tilde{u}$ values (vanishing $ U^{\tilde{u}}$) parameterize outgoing trajectories, while 
constant $\tilde{v}$ values parameterize ingoing trajectories.

\subsection*{Relationships between the locally-flat and accelerating conformal coordinates}
 \indent \indent
We next examine the relationship between the outgoing
conformal coordinate $u$ of the accelerating system
with that $\mathcal{U}$ of the locally-flat freely-falling
frame momentarily at rest with the accelerating system. 
As will be motivated in the Appendix, this relationship
is needed to describe the spectral
nature (i.e. frequency distribution) of outgoing massless radiations.
By direct substitution into the form (\ref{Eq:uFvF}),
\begin{equation}
\Delta \tilde{u}  \simeq \frac{\Delta \tilde{\mathcal{U}}_\text{moving}}{1 - \sqrt{\frac{R_S}{r_p} } } +
\frac{R_S}{4 r_p^2} \: \frac{\Delta \tilde{\mathcal{U}}_\text{moving}^2}{ \left( 1 - \sqrt{\frac{R_S}{r_p} }  \right)^2}
+ \mathcal{O}[\Delta \tilde{\mathcal{U}}_\text{moving}^3] ,
\end{equation}
where $\Delta \tilde{\mathcal{U}}_\text{moving} \equiv
(Y^0-\omega_p)-(Y^1-\rho_p)$
in terms of the coordinates in equations (\ref{Eq:omgrhoYYFFF})
and (\ref{Eq:ctrYYFFF}).  A simple Lorentz transformation given by
$\Delta \tilde{\mathcal{U}}_\text{moving} = \sqrt{\frac{1+\beta_Y}{1-\beta_Y}} \: \Delta \mathcal{U}$
relates the momentarily stationary conformal coordinate
$\Delta \mathcal{U}$ to $\Delta \tilde{\mathcal{U}}_\text{moving}$,
where $\beta_Y = -\sqrt{\frac{R_S}{r_p}}$. 
Thus, the conformal parameters are related by
\begin{equation}
\Delta \tilde{u} \simeq \frac{\Delta \mathcal{U}}{\sqrt{1-\frac{R_S}{r_p}}}+\frac{1}{2}\left(\frac{R_S}{2r_p}\right) \left(\frac{\Delta \mathcal{U}}{\sqrt{1-\frac{R_S}{r_p^2}}} \right)^2 ,
\label{Eq:DeltaUtilde}
\end{equation}
good to second order in deviations of the coordinate from
the point of measurement. 

Examining the expressions in (\ref{Eq:NullForms}) allows the metric to be re-expressed
in the forms
\begin{equation}
ds^2=  - \left[ \left( 1 - \sqrt{ \frac{R_S}{r} }  \right) d\tilde{u}\right] \: \left[ \left( 1 + \sqrt{ \frac{R_S}{r} }  \right) d\tilde{v} \right]
= -du \, dv \simeq -d\mathcal{U} \: d\mathcal{V}.
\end{equation}
As defined, the conformal coordinates $(u,v)$ and $(\mathcal{U},\mathcal{V})$ are guaranteed to be quantified
using the same units and measuring sticks. 
From the form of $d\tilde{u}=d(ct-r_{\tilde{u}})$, it is clear from the time parameter that this represents a form of
the conformal coordinate at rest in the freely-falling $(\omega=ct, \rho)$ coordinate frame.  This means that in the vicinity of the
observation point, the boosted conformal coordinate satisfies
\be
\Delta u \simeq \left( 1-\sqrt{\frac{R_S}{r_p}}  \right) \Delta \tilde{u} \simeq 
\left( 1-\sqrt{\frac{R_S}{r_p}}  \right)  \sqrt{\frac{ 1+\sqrt{\frac{R_S}{r_p}}}{ 1-\sqrt{\frac{R_S}{r_p}}}} \: \Delta \tilde{u}  
\simeq \sqrt{1-\frac{R_S}{r_p}} \: \Delta \tilde{u},
\ee
(which can be more directly obtained using $d\tau=\sqrt{1-\frac{R_S}{r}} \: dt$),
and similarly for $\Delta v$.  
Substitution of this result into the equation (\ref{Eq:DeltaUtilde}) defines the necessary form of the relation
between the conformal coordinates consistent with the principle of equivalence:
\begin{equation}
\Delta u \simeq \Delta \mathcal{U} + 
\frac{1}{2} \frac{\left(\frac{R_S}{2r_p^2} \right)}{\sqrt{1-\frac{R_S}{r_p}}}\left(\Delta \mathcal{U} \right)^2 .
\label{uUeqn}
\end{equation}
Algebraically defining an acceleration $\bar{a}$ of the form
\begin{equation}
\frac{\bar{a}}{c^2} \equiv \frac{\left(\frac{R_S}{2r_p^2} \right)}{\sqrt{1-\frac{R_S}{r_p}}}
\label{Eq:abar}
\end{equation}
provides the second-order expression that the resulting dimensionless functions must satisfy:
\begin{equation}
\frac{\bar{a}}{c^2}\Delta u \simeq \frac{\bar{a}}{c^2}\Delta \mathcal{U}+ \frac{1}{2}\left(\frac{\bar{a}}{c^2}\Delta \mathcal{U} \right)^2 .
\label{uUaeqn}
\end{equation}
It only remains to physically interpret the algebraic parameter $\bar{a}$.


\subsection*{Local proper acceleration}
\indent \indent
We next demonstrate the form of the proper acceleration of the curvilinear observers.  In flat space-time,
 the time-like four-velocity ($\vec{U} \equiv \frac{\vec{dY}}{dc\tau}$) invariant $\vec{U}\cdot\vec{U}=-1$ 
can be differentiated to demonstrate its orthogonality to
the four-acceleration $\vec{A} \equiv \frac{\vec{dU}}{dc\tau}$, requiring $\vec{A}$ to be space-like.
This implies that the proper-acceleration $a$ defined in the rest
frame of an accelerating observer satisfies $\vec{A} \cdot \vec{A}=\frac{a_{proper}^2}{c^2}$. 
Affine connections $ \Gamma^\alpha_{\beta \lambda}$ can then be defined via the
principle of equivalence relating descriptions of accelerations in a locally-flat freely-falling
frame to descriptions of those accelerations using curvilinear coordinates via
\be
\frac{\tilde{a}^\mu}{c^2}=\frac{\partial \tilde{x}^\mu}{\partial x^\alpha} \left [
\frac{d^2 x^\alpha}{d c\tau^2} + \Gamma^\alpha_{\beta \lambda}
\frac{dx^\beta}{dc\tau} \frac{dx^\lambda}{dc\tau}
\right ] \quad \Rightarrow \quad \left( a_{proper} \right )^2 =
\tilde{g}_{\mu \nu} \tilde{a}^\mu \tilde{a}^\nu .
\label{Eq:ProperAccel}
\ee
Using the relationships between $(ct,r)$ and  the freely falling coordinates 
$(\omega,\rho)$ from
(\ref{Eq:LamaitreJL}) in equation (\ref{Eq:ProperAccel}) for observers with fixed $r$
($\frac{dr}{dc\tau}=0$) results in the form
\be
\frac{a_{proper}}{c^2}=\frac{ \frac{R_S}{2 r^2}}{\sqrt{1-\frac{R_S}{r} } }.
\label{Eq:RivAccel}
\ee
One alternatively can use the $(Y^0,Y^1)$
locally flat coordinates (\ref{Eq:ctrYYFFF}) to determine
the proper acceleration of the point $(ct_p,r_p)$. 
From the expressions in (\ref{Eq:Yderivs}), the form in (\ref{Eq:RivAccel}) is locally verified.

Therefore, we see that the algebraically defined dimensional acceleration
from expression (\ref{uUaeqn}) 
is the same as the \emph{local} proper acceleration of the
stationary observer using the curvilinear coordinates,
$\bar{a}=a_{proper}$.
This means that
any physical effects implemented into the curved space-time via the principle of equivalence
\emph{anywhere} within the (exterior) geometry necessarily involve only \emph{local} proper accelerations.


\subsection*{Flatness scales}
\indent \indent
To insure \emph{local flatness}, the locally freely falling frame momentarily at rest with the accelerating frame satisfying Eqn. (\ref{uUeqn})
is only valid for coordinate variations small compared to lengths associated with curvatures:
 \be
 \Delta \mathcal{U} << \sqrt{ \frac{r_p^3} {2 R_S} } .
\label{LFcondition}
\ee
This means that any micro-physical measurement scales must satisfy this criterion. 
Likewise, the form in Eqn. (\ref{uUeqn}) can reliably represent an expansion of a function only if
\be
\left [ \frac{\left(\frac{R_S}{2r_p^2} \right)}{\sqrt{1-\frac{R_S}{r_p}}} \right ] \Delta \mathcal{U}  <<1 .
\ee
These relations imply that any functional form can be reliably tested using the principle of equivalence for
$r_p\gapproxeq \frac{9}{8} R_S$.  For local measurements nearer than this, testable functional dependencies are not reliable. 
Furthermore, an expression of the form $y \simeq x+ \frac{1}{2}x^2$ as in equation (\ref{uUaeqn}) could
represent $y=-log(1+x)$ (which results in a ``thermal" distribution),
the geometric series $y=\frac{x}{1-\frac{x}{2}}$, or numerous alternative
functional forms.  What \emph{has} been established is the use of the \emph{local} proper acceleration
in the power series expansion of the dimensionless variables that represents the functional form.


\section{Comparisons of Accelerating Observers}
\indent \indent
In this section we will compare descriptions of accelerating observers in various geometries. 
Observers in flat space-time who undergo a constant local proper acceleration
can construct a (Rindler) coordinate patch $(\omega_a,\rho_a)$ that relates to the
global Minkowski coordinates $(\xi^0,\xi^1)$ via
\be
\xi^0 (\omega_a,\rho_a)= \rho_a \sinh\left(\frac{a \, \omega_a }{c^2} \right) \quad \textnormal{and} \quad
\xi^1 (\omega_a,\rho_a) = \rho_a \cosh\left(\frac{a \,\omega_a }{c^2}\right).  
\label{Eq:RindlerCoords}
\ee
The outgoing horizon $\xi^0-\xi^1=0$, which is an asymptote of the accelerating trajectory, has coordinate $\rho_a=0$. 
Furthermore, Rindler coordinates $(ct,x) \equiv (\omega_a, \: \frac{c^2}{a} log\left( \frac{a \, \rho_a}{c^2} \right))$
conducive for developing conformal coordinates can be utilized,
resulting in the following forms for the metric:
\be
ds^2 = -(d \xi^0)^2 + (d \xi^1)^2 = - \rho_a^2 \: d\omega_a^2 + d\rho_a^2
= e^{2 \frac{a \, x}{c^2}} \, \left( -dct^2 + dx^2 \right).
\label{Eq:Rindlermetric}
\ee
From this expression, one can directly construct conformal inertial coordinates
$(\mathcal{U}\equiv \xi^0-\xi^1, \mathcal{V} \equiv \xi^0+\xi^1)$ and
accelerating coordinates $(\tilde{u} \equiv ct-x,\tilde{v} \equiv ct+x)$. 
The coordinate $x$ maps the horizon $\rho_a=0$ to $x_H=-\infty$.

The near-horizon form of the 1+1 dimensional Schwarzschild metric 
\begin{equation}
ds^2= -\left(1-\frac{R_S}{r}\right)c^2dt_s^2 + \frac{dr^2}{1-\frac{R_S}{r}}
\label{Schwarzschildmetric}
\end{equation}
shares a coordinate singularity (with $t_S \rightarrow \infty$) at the horizon $r_H=R_S$
with the Rindler metric at $\rho_{aH}=0$.  The \emph{near-horizon} identifications
$\rho_a \simeq 2 \sqrt{R_S (r-R_S)}$ and $d\omega_a \simeq \frac{dct_S}{2 R_S}$
for $\left( \frac{r-R_S}{R_S} \right) <<1$
are used to assert an ``effective acceleration" $a_{effective}=\frac{c^2}{2 R_S}$
resulting in a thermal temperature associated with the horizon by asymptotic Schwarzschild
observers who use $t_S$ as the time parameter conjugate to their quantum energy measurements,
as will be motivated in the Appendix.


\subsection*{Near-horizon Painlev\'e-Gullstrand}
\indent \indent
Next, we examine the near-horizon form of the Painlev\'e-Gullstrand metric.  We will define
$\Delta \equiv \frac {r-R_S}{R_S}$, and examine local analytic coordinate transformation
to a new time coordinate $\tilde{ct}$:
\begin{equation}\label{RindCorres}
dct \equiv d\tilde{ct} +  R_S \frac{d\Delta}{\Delta},
 \text{with proper distance } d\tilde{\rho}= R_S \frac{d\Delta}{\sqrt{\Delta}} .
\end{equation}
The metric (\ref{RivIntervalctr}) transforms into a diagonal form given by
\begin{equation}\label{NearHorizon}
ds^2=  \frac{R_S^2}{\Delta}\left( - \frac{\Delta^2}{R_S^2} \, d\tilde{ct}^2  +
d\Delta^2 \right) =
-\left ( \frac{\tilde{\rho}}{2 R_S} \right)^2 \: d\tilde{ct}^2 + d \tilde{\rho}^2 .
\end{equation}
It is clear that the transformation is not analytic at the horizon $\Delta \rightarrow 0$.  This is expected since
no analytic transformation can take non-singular coordinates into singular coordinates. 
This \emph{local} transformation can be utilized by even distant observers, but
\emph{asymptotic} Schwarzschild observers require a singular transformation
on (\ref{RivIntervalctr}) in order to identify $ct_S$ as $\tilde{ct}$.

\subsection*{Probes of inertial and accelerating ground states}
\indent \indent
As will be motivated in the Appendix, inertial probing of accelerating ground states,
as well as probing of inertial ground states by accelerating probes, involve representing
the quantum operators of one system in terms of the other's using canonical
transformations.  For outgoing massless radiations, the formulations requires
a calculation of the inertial conformal coordinate in terms of the accelerating
conformal coordinate, $\mathcal{U}(u)$.

For a Rindler observer undergoing constant acceleration in Minkowski space-time,
the needed relations
$(\mathcal{U} = \xi^0-\xi^1, \mathcal{V}= \xi^0+\xi^1)$ and
$(\tilde{u} = ct-x,\tilde{v}=ct+x)$
can be calculated from (\ref{Eq:RindlerCoords}) with
$\rho_a = \frac{c^2}{a} e^{\frac{a}{c^2} x}$, yielding
\be
\mathcal{U}(\tilde{u})=-\frac{c^2}{a} \: e^{- \frac{a}{c^2} \tilde{u}} \quad \textnormal{and}\quad
\mathcal{V}(\tilde{v})=\frac{c^2}{a} \: e^{ \frac{a}{c^2} \tilde{v}} .
\label{Eq:RindlerConformalRelations}
\ee
The form $\mathcal{X}=-e^{-x}$ is unique in that rescaling the inertial coordinate $\mathcal{X} \rightarrow \lambda \mathcal{X}$
simply results in a shift of the accelerating coordinate $ \lambda \mathcal{X} = -e^{-(x-log \lambda)}$. This can be interpreted for
uniform acceleration as implying that the rescaling of inertial light-like trajectories is equivalent to a Lorentz (Doppler) transformation
due to shifting the onset of acceleration.

For near horizon, as well as asymptotic Schwarzschild observers in 1+1 dimensions, we expect relations analogous to
(\ref{Eq:RindlerConformalRelations}) with a \emph{finite} effective acceleration to be valid.  For Painlev\'e-Gullstrand observers, the
result obtained using the principle of equivalence (\ref{uUaeqn}) is likewise consistent with a global form
$\Delta \mathcal{U}(\Delta u)=-\frac{c^2}{a_{proper}} \: e^{- \frac{a_{proper}}{c^2} \: \Delta u}$, only with
effective accelerations that becomes \emph{vanishingly} smaller for more distant observers.

\section{Discussion and Conclusions}
\indent \indent 
In this paper, the principle of equivalence was used to examine macro- and micro-physical measurements
using the  Painlev\'e-Gullstrand metric. 
To do this, a global set of freely falling coordinates utilizing the proper time of these fallers
as the non-singular global temporal coordinate was developed. 
Since these proper times represent observable clocks, one is expected to be able to covariantly examine
quantum dynamics using conjugate energies with minimal interpretation.
From these global coordinates, locally-flat
freely-falling coordinates that incorporate curvature
(and generally, local energy density) scales were obtained.

Un-ambiguous algebraic relationships were established between the coordinates
(including conformal coordinates) of
stationary accelerating observers with fixed coordinate $r=r_p$ and those of
the freely-falling observers that are momentarily at rest with the stationary observers, allowing
the mutual probing of local phenomena.  In particular, the relationships demonstrate that
the conformal coordinates used by the two sets of observers differ only through
the proper acceleration of the $(ct_p,r_p)$ observers, without Doppler shift. 

Phenomena associated with the local proper accelerations of stationary
observers were found to be \emph{consistent} with those experienced by Rindler observers
\emph{away} from the horizon in regions of sufficient flatness. 
This means that \emph{globally}, any local radiations detected by observers with static curvilinear coordinates
in this geometry are related \emph{only} to local proper accelerations. 
We expect increasingly distant stationary observers to measure
vanishingly smaller phenomena associated with their proper accelerations,
decreasing to zero for very distant observers in contrast
to what is expected for asymptotic Schwarzschild observers.  The results of
measurements of accelerating probes on inertial ground states are
found to be consistent with those done by observers in Rindler space.

\renewcommand{\theequation}{A.\arabic{equation}}
\setcounter{equation}{0}

\pagestyle{empty}
\section*{Appendix}
\indent \indent
For completeness, the spectral distribution of radiations from the ground states
of others as measured by inertial and accelerating observers will be briefly sketched. 
A more complete exposition is available in the references\cite{Hawking, Mukhanov}.

A massless scalar field $\hat{\phi}(ct,x)$ can be expressed using operators $\hat{b}(k)$ quantized in the laboratory frame of reference
$(ct,x)$ in the form
\begin{equation}
\hat{\phi}(ct,x)= \sqrt{\frac{\hbar}{2\pi}} \int_{-\infty}^{\infty} dk \sqrt{\frac{1}{2|k|}}\left[ e^{i \left( -|k|ct+kx \right)} \hat{b}(k)+ 
e^{-i\left(-|k|ct+kx \right)} \hat{b}^\dagger(k) \right].
\label{Eq:GenQuantumField}
\end{equation}
The operators  $\hat{b}(k)$ satisfy canonical commutation relations as a function of wave-number $k$, and are assumed to
annihilate the ground state of radiations in the frame $(ct,x)$.  Assuming that conformal coordinates can be directly expressed
using $u=ct-x$ and $v=ct+x$, the field can be written (with $\hat{b}\left( \frac{\omega}{c} \right)\equiv \hat{b}_\omega$) as 
\begin{align}
\hat{\phi}(u,v) 
 &= \sqrt{\frac{\hbar}{2 \pi}}\int_0^\infty d\omega \sqrt{\frac{1}{2 \omega c}} \left[ e^{-i\frac{\omega}{c}u} \hat{b}_\omega
+ e^{i\frac{\omega}{c}u} \hat{b}^\dagger_\omega+
e^{-i\frac{\omega}{c}v} \hat{b}_{-\omega} + e^{i\frac{\omega}{c}v} \hat{b}^\dagger_{-\omega} \right] \nonumber \\
&\equiv \: \hat{\phi}_+(u)+\hat{\phi}_-(v) , \label{uvquant}
\end{align}
where the dispersion relation $\omega= c|k|$ for \emph{positive} frequencies has been utilized.

In a similar way, the scalar field can be expressed using operators quantized in the freely falling
or inertial frame $(Y^0,Y^1)$ as:
\be
\hat{\phi}(u,v) = \hat{\Phi}(\mathcal{U},\mathcal{V})  \equiv \: \hat{\Phi}_+ (\mathcal{U}) + \hat{\Phi}_-(\mathcal{V}) , 
\label{UVquant}
\ee
where operators $\hat{B}(K)$ expressed in terms of inertially defined frequencies replace the operators
$\hat{b}(k)$ in equation (\ref{uvquant}).  A canonical Bogoliubov transformation can be found to
connect the operators $\left( \hat{b}(k),\hat{b}^\dagger(k) \right)$ to
$\left( \hat{B}(K),\hat{B}^\dagger(K) \right)$, preserving the commutation relationships. 
Since outgoing/ingoing light-like trajectories must be represented by constant $(u,\mathcal{U})/(v,\mathcal{V})$
for either observer, one can conclude that 
$\mathcal{U}=\mathcal{U}(u)$ and $\mathcal{V}=\mathcal{V}(v)$, i.e. the transformations do not mix
ingoing with outgoing fields in (\ref{UVquant}).  Thus
$\hat{\Phi}_+ (\mathcal{U}(u)) = \hat{\phi}_+(u)$ and $\hat{\Phi}_-(\mathcal{V}(v)) = \hat{\phi}_-(v)$.

Explicitly, for outgoing massless radiations
\begin{align}
\hat{\Phi}_+(\mathcal{U}(u)) &= \hat{\phi}_+(u)  \label{Eq:PhiEqualsphi} \\
\int_0^\infty d\Omega \frac{1}{\sqrt{2 \Omega c}} \left[ e^{-i\frac{\Omega}{c} \mathcal{U}} \hat{B}_\Omega
+ e^{i\frac{\Omega}{c} \mathcal{U}} \hat{B}^\dagger_\Omega \right] 
&= \int_0^\infty d\omega \frac{1}{\sqrt{2 \omega c}} \left[ e^{-i\frac{\omega}{c}u} \hat{b}_\omega
+ e^{i\frac{\omega}{c}u} \hat{b}^\dagger_\omega \right] . \nonumber 
\end{align}
Recall that, as was done in (\ref{uvquant}), Fourier transforms of fields $\hat{\Psi}(u)$ over all wave-numbers $k$ can be alternatively
expressed in terms of integrals over positive frequencies $\omega$ only, generally requiring that the Fourier transformed fields $\tilde{\hat{\Psi}}(k)$
satisfy 
\be
\hat{\Psi}(u)= \int_{-\infty}^\infty \frac{d k}{\sqrt{2\pi}} \tilde{\hat{\Psi}}(k) e^{-i k u} =
\int_0^\infty \frac{d\left(\frac{\omega}{c}\right)}{\sqrt{2\pi}}\left[\tilde{\hat{\Psi}}\left(-\frac{\omega}{c}\right)e^{\frac{i}{c}\omega u} + 
\tilde{\hat{\Psi}}\left(\frac{\omega}{c}\right) e^{-\frac{i}{c}\omega u} \right] .
\label{Eq:GenFourierTransfs}
\ee
Thus, we can make the following identification for the Fourier transform:
\begin{equation}
\tilde{\hat{\phi}}_+(k) =
\left\{
\begin{array}{ll}
\sqrt{\frac{\hbar}{2\omega c}} \: \hat{b}_\omega
&\text{for } k>0 , \\
\sqrt{\frac{\hbar}{2 \omega c}} \: \hat{b}^\dagger_\omega
&\text{for } k<0 .
\end{array}
\right.
\label{Eq:Smallb}
\end{equation}

Again utilizing (\ref{Eq:GenFourierTransfs}),
the Fourier transform of $\hat{\Phi}_+(\mathcal{U}(u))$ results in the expression
\begin{align}
\tilde{\hat{\phi}}_+(k) &=\int_{-\infty}^{\infty} \frac{du}{\sqrt{2\pi}} \, e^{i\frac{\omega}{c}u} \,
\sqrt{\frac{\hbar}{2 \pi}}\int_0^\infty d\Omega \frac{1}{\sqrt{2 \Omega c}} \left(e^{-i\frac{\Omega}{c} \mathcal{U}}\hat{B}_\Omega +
e^{i \frac{\Omega}{c} \mathcal{U}}\hat{B}^\dagger _\Omega \right) \nonumber \\
&= \int_{0}^{\infty} d\Omega \: \sqrt{\frac{\hbar}{2\Omega c}} \left[F(\Omega,\omega) \: \hat{B}_\Omega +
F(-\Omega,\omega) \: \hat{B}^\dagger_\Omega \right],
\end{align}
where
\begin{equation}
F(\Omega,\omega)\equiv \int_{-\infty}^\infty \frac{du}{2\pi} \: e^{\frac{i}{c}(\omega u -\Omega \, \mathcal{U}(u))}.
\label{Eq:Ffunction}
\end{equation}
From (\ref{Eq:Smallb}), we obtain the following relations between the canonical operators:
\begin{equation}
\hat{b}_\omega = \int_0^\infty d\Omega \left[\sqrt{\frac{\omega}{\Omega}}F(\Omega,\omega)\hat{B}_\Omega +
\sqrt{\frac{\omega}{\Omega}}F(-\Omega,\omega)\hat{B}^\dagger_\Omega \right].
\label{Eq:OperatorTransforms}
\end{equation}
This general equation relates canonical massless scalar operators
that act upon accelerating and inertial ground states used by the respective observers.

In particular, we demonstrate the spectral characteristics of outgoing massless radiations from an
inertial Minkowski ground state as measured by a Rindler probe
with proper acceleration $a$. 
Substituting (\ref{Eq:RindlerConformalRelations}) to relate inertial and
accelerating conformal coordinates, the auxiliary function (\ref{Eq:Ffunction}) takes the form
\begin{equation}
F_{R}(\Omega,\omega)\equiv \int_{-\infty}^\infty \frac{d \tilde{u}}{2\pi}
e^{\frac{i}{c}\left(\omega \, \tilde{u} +\Omega \, \frac{c^2}{a} \, e^{-\frac{a \tilde{u}}{c^2}}\right)}.
\label{Eq:RindlerAuxiliaryFcn}
\end{equation}

Calculating the Minkowski ground state
expectation value of the accelerating number operator of mode frequency $\omega$
gives the number of excitations of this mode in the Minkowski ground state $\left| \, 0_M \right \rangle$:
\begin{align}
\langle N \rangle_\omega=\left\langle 0_M \left| \hat{b}^\dagger_\omega \hat{b}_\omega \right|0_M \right\rangle=
\int_0^\infty d\Omega \left|\sqrt{\frac{\omega}{\Omega}}F_R(-\Omega,\omega) \right|^2 .
\label{Eq:GenNumExpectVal}
\end{align}
The auxiliary function (\ref{Eq:RindlerAuxiliaryFcn}) for Rindler observers can be expressed in terms of a gamma function:
\begin{equation} \label{reln}
F_R(\Omega,\omega)=\frac{c^2}{2\pi a}e^{\left(\frac{ic\omega}{a} \right)\ln\left(- \frac{i c \Omega}{a} \right)}\Gamma\left(-\frac{ic\omega}{a} \right) =e^{\frac{\pi c \omega}{a}}F_R(-\Omega,\omega) .
\end{equation} 
Furthermore, the canonical commutation relation $\left[ \hat{b}_\omega, \hat{b}^\dagger_{\omega'} \right]$ gives
\begin{align}\label{bcomrel}
&\left[\hat{b}\left( \frac{\omega}{c}\right),\hat{b}^\dagger\left( \frac{\omega'}{c}\right)\right]= \delta\left(\frac{\omega}{c}-\frac{\omega'}{c} \right) \nonumber\\
&= \int_0^\infty d\Omega \left( \sqrt{\frac{\omega'}{\Omega}}F_R^*(\Omega,\omega')\sqrt{\frac{\omega}{\Omega}}F_R(\Omega,\omega)-\sqrt{\frac{\omega'}{\Omega}}F_R^*(-\Omega,\omega')\sqrt{\frac{\omega}{\Omega}}F_R(-\Omega,\omega) \right) .
\end{align}
Substituting (\ref{reln}) into (\ref{bcomrel}), and setting $\omega = \omega'$, we get:
\begin{equation}
\delta(0)= \left(e^{\frac{2\pi c \omega}{a}}-1 \right)\int_0^\infty d\omega \left|\sqrt{\frac{\omega}{\Omega}}F_R(-\Omega,\omega) \right|^2 
\stacksymbols{\Rightarrow}{(\ref{Eq:GenNumExpectVal}) }{8}{0.5} \langle N_\omega \rangle= \frac{\delta(0)}{\left(e^{\frac{2\pi c \omega}{a}}-1 \right)} .
\label{Eq:RindlerNumExpectVal}
\end{equation} 

Using the common $L\rightarrow \infty$ identification $\frac{L}{2 \pi} \delta_{k_r', k_s} \doteq \delta(k'-k)$
in the result from(\ref{Eq:RindlerNumExpectVal}),
the number density in 1+1 dimensions is effectively identified as
$\langle n \rangle_\omega \doteq \frac{1}{2\pi \delta(0)}\langle N \rangle_\omega$.
yielding
\begin{equation}
\langle n \rangle_\omega \doteq \frac{1}{e^{\frac{2\pi c}{a} \,  \omega}-1}.
\end{equation}
This is precisely the Planck form of black body radiation, if the effective temperature is
given by
\begin{equation}
T=\frac{\hbar \, a}{2\pi c k_B}.
\end{equation}


\end{document}